\documentclass[twocolumn, amssymb, pra, ,showpacs, superscriptaddress, nobibnotes]{revtex4-1}

\usepackage{graphicx}% Include figure files
\usepackage{dcolumn}% Align table columns on decimal point
\usepackage{bm}% bold math

\begin{document}

\title{Index of refraction of molecular nitrogen for sodium matter waves}

\author{J. Loreau}
\thanks{Present address: Service de Chimie Quantique et Photophysique, Universit\'e Libre de Bruxelles, Belgium.}
\affiliation{Institute for Theoretical Atomic, Molecular and Optical Physics, Harvard-Smithsonian Center for Astrophysics, Cambridge, Massachusetts 02138, USA}
\author{V. Kharchenko}
\affiliation{Institute for Theoretical Atomic, Molecular and Optical Physics, Harvard-Smithsonian Center for Astrophysics, Cambridge, Massachusetts 02138, USA}
\affiliation{ Physics Department, University of Connecticut, 2152 Hillside Road, Storrs, CT 06269, USA }
\author{A. Dalgarno}
\affiliation{Institute for Theoretical Atomic, Molecular and Optical Physics, Harvard-Smithsonian Center for Astrophysics, Cambridge, Massachusetts 02138, USA}

%\date{\today}

\begin{abstract}

We calculate the index of refraction of sodium matter waves propagating through a gas of nitrogen molecules. 
We use a recent {\it ab initio} potential for the ground state of the NaN$_2$ Van der Waals complex to perform quantal close-coupling calculations and compute the index of refraction as a function of the projectile velocity. We obtain good agreement with the available experimental data. 
We show that the refractive index contains glory oscillations, but that they are damped by the averaging over the thermal motion of the N$_2$ molecules. These oscillations appear at lower temperatures and projectile velocity. 
We also investigate the behavior of the refractive index at low temperature and low projectile velocity to show its dependence on the rotational state of N$_2$, and discuss the advantage of using diatomic molecules as projectiles. 

\end{abstract}


\maketitle
%\tableofcontents

\section{Introduction}

The refractive index is usually used to describe the propagation of light through a medium, but it can also be applied to the propagation of matter waves through a dilute gas. In that case, the refractive index is a complex quantity that depends on collisions between the particles of the propagating wave and the gas \cite{Lax1951}. 
The refractive index can be measured using atom or molecule interferometry and can help to determine properties of atom-atom or atom-molecule interaction potentials \cite{Berman1997}. 
It is proportional to the complex forward scattering amplitude $f(k,0)$. The imaginary part describes the attenuation of the beam and is related to the total scattering cross section, while the real part is associated with the phase shift due to the interaction between projectile  particles and atoms or molecules of the buffer gas.

The first experiment on matter wave interferometry was realized with sodium atoms propagating in gas cells containing various gases \cite{Schmiedmayer1995}. Subsequent experiments led to the observation of glory oscillations in the index of refraction as a function of the beam velocity for sodium matter waves in a medium of noble gases or nitrogen \cite{Roberts2002}, and to the investigation of Na$_2$ matter waves \cite{Chapman1995}. These works motivated studies on the theory of the refractive index for atoms \cite{Vigue1995,Kharchenko2001b,Forrey2002,Champenois2008}, which were used in combination with accurate interaction potentials to calculate the refractive index of various noble gases as a function of the sodium beam velocity \cite{Audouard1995,Forrey1996,Forrey1997,Blanchard2003}. A comparison between the experimental data and the theoretical calculations can be found in Ref. \cite{Cronin2009}.
More recently, the refractive index of noble gases for lithium matter waves was measured \cite{Jacquey2007} and investigated theoretically \cite{Lemeshko2010,Bjorgen2011}. 

Theoretical analysis of the matter wave refraction has  been carried out mostly for atomic projectiles and bath gases. Collisions of molecular species may involve transitions between internal ro-vibrational degrees of freedom of a projectile or/and bath gas particles and the computation of the refractive index becomes a significant problem.
In this work, we compute the index of refraction for sodium matter waves propagating through a gas of molecular nitrogen N$_2$. This system, for which experimental data are available over a wide range of projectile velocities \cite{Roberts2002}, has not been investigated theoretically. A comparison with the experiment also provides a check of the quality of the {\it ab initio} potential for the NaN$_2$ complex.
In Sec. \ref{section_theory}, we recall the definition of the refractive index in terms of the scattering amplitude, and we detail the calculation of the amplitude for atom-molecule collisions.
In Sec. \ref{section_results}, we compare our calculations to the experimental data and discuss the dependence of the scattering amplitude on the rotational level at low temperature and projectile velocity.

\section{Theoretical methods}\label{section_theory}

\subsection{The refractive index}

In an interferometry experiment, a beam of projectile atoms with mass $m_p$ and velocity $v_p$ propagates through a cell containing a gas of target particles of density $n_t$ and mass $m_t$. 
This gas can be described as an effective medium with a complex refractive index that depends on the scattering amplitude.
Two expressions for the refractive index have been used in the literature. The first was derived by Champenois {\it et al.} \cite{Champenois2008},
\begin{equation}\label{index_Champenois}
n=1+2\pi n_t \frac{m_p+m_t}{m_t}\frac{1}{k^2_p} \left\langle f(k_r) \right\rangle \ ,
\end{equation}
while Forrey {\it et al.} wrote the refractive index as \cite{Forrey1996}
\begin{equation}\label{refractive_index1}
n=1+2\pi n_t \frac{1}{k_p} \left\langle \frac{f(k_r)}{k_r} \right\rangle
\end{equation}
In Eqs. (\ref{index_Champenois}) and (\ref{refractive_index1}), $f(k_r)=f(k_r,\theta=0)$ is the forward scattering amplitude calculated in the center of mass frame, $k_r$ is the wave number for the relative motion of the colliding particles, $\hbar k_r=\mu v_r $, $\mu$ is the reduced mass of the system, and $k_p$ is the wave number of the projectile particle in the laboratory frame, $\hbar k_p=m_p v_p$. The brackets $\langle \cdot \rangle$ denote a thermal averaging over the distribution of target atom velocities. The real part of the refractive index is related to the phase shift accumulated by the matter wave due to scattering with the gas, while the imaginary part corresponds to the attenuation of the beam. 
All the results presented in this work were calculated using the expression (\ref{refractive_index1}). Eqs. (\ref{index_Champenois}) and (\ref{refractive_index1}) can yield very different results at low projectile velocities, as was discussed by Champenois {\it et al.} \cite{Champenois2008}. However, for the system under consideration we found that the discrepancy between the two formulas was small even at low projectile velocities.

While the real and imaginary parts of the refractive index can in some cases be measured accurately \cite{Jacquey2007}, it is useful to define the quantity $\rho$ as the ratio of the phase shift to the attenuation,
\begin{equation}\label{rho_def}
\rho(v_p)=\frac{\textrm{Re}(n-1)}{\textrm{Im}(n)} \ .
\end{equation}
The main advantage of measuring $\rho$ instead of $n$ is that it is independent of the gas density \cite{Schmiedmayer1995}.

\subsection{Thermal averaging}
While the projectile atoms have a definite velocity, this is not the case for the target atoms and a thermal averaging must be carried out.
The thermal distribution for the velocity of the target atoms or molecules in the laboratory frame ${\bf v}_t$ is given by a Maxwellian function:
\begin{equation} \label{MB_distrib}
P({\bf v}_t)
= \left( \frac{\beta}{\sqrt{\pi}} \right)^{3} \exp\left(- \beta^2 {\bf v}_t^2 \right) \ , \qquad \beta^2=\frac{m_t}{2k_BT}
\end{equation}
where $T$ is the temperature of the target gas. In the experiment \cite{Roberts2002}, $T=300$ K. However, the scattering amplitude is calculated in the center of mass frame, so that it is necessary to express this distribution as a function of the relative velocity ${\bf v_r=v_p-v_t}$. After integration over the angular part, the distribution (\ref{MB_distrib}) becomes
\begin{equation}\label{MB_distrib2}
P(v_r,v_p) =\frac{2\beta}{\sqrt{\pi}} \; \frac{v_r}{v_p} \; \exp\left(- \beta^2 (v_p^2+ v_r^2) \right) \sinh\left( 2\beta^2v_p v_r \right)  
\end{equation}
and the averaging of the scattering amplitude is given by
\begin{equation}\label{average_ampl}
\langle f(k_r)\rangle = \int_0^\infty f(k_r) P(v_r,v_p) dv_r
\end{equation}

\subsection{Calculation of the scattering amplitude}

We calculated the forward scattering amplitude using the quantum close-coupling method  \cite{Arthurs1960} while treating N$_2$ as a rigid rotor. The experimental temperature is much lower than the vibrational spacing of N$_2$ so that all collisions occur in the ground state of vibrational motion $v=0$. The probability of excitation of N$_2$ to higher vibrational levels is small at the experimental collision velocities and these channels may be ignored in calculations of the refractive index.
The rotational energies of N$_2$ were obtained using the rotational constant \cite{Bendtsen1974,Lothus1977} $B_e=1.99824$ cm$^{-1}$.

In the close-coupling method, the wave function is expanded in terms of radial and rotational functions. This expansion is then inserted into the Schr\"odinger equation, leading to a set of second order equations in which the couplings are represented by matrix elements of the intermolecular potential $V(R,\theta)$  \cite{Arthurs1960,Flower2007}
. The close-coupled (CC) equations take a simple form if  the total angular momentum ${\bf J = j+l}$ (where ${\bf j}$ is the angular momentum of N$_2$ and ${\bf l}$ is the orbital momentum of the collision) is introduced. The total angular momentum is conserved during the collision, so that the coupled equations are block-diagonal in $J$. The $S$ matrix elements $S^J_{j^\prime l^\prime jl}$ are obtained from the solution of the CC equations, and the scattering amplitude for the transition from an initial rotational state $j$ with projection $m_j$ to a final rotational state $j^\prime$ with projection $m_{j^\prime}$ is given in terms of the $S$ matrix elements as \cite{Arthurs1960}
\begin{eqnarray}  \label{scatt_ampl_theta}
f_{jm_j\rightarrow j^\prime m_{j^\prime}}(E_r,\theta) && = 
\frac{\sqrt{\pi}}{k_j} \sum_l \sum_{l^\prime m_{l^\prime}} i^{l-l^\prime} \sqrt{2l+1} \ Y_{l^\prime m_{l^\prime}}(\theta) \nonumber \\
&& \times \sum_{J} (-)^{l+l^\prime+j+j^\prime} (2J+1) 
\left(\begin{array}{ccc} j & l & J \\ m_j & 0 & -M\end{array}\right) \nonumber \\ 
&& \times \left(\begin{array}{ccc} j^\prime & l^\prime & J \\ m_{j^\prime} & m_{l^\prime} & -M\end{array}\right) 
\vert \delta_{ll^\prime}\delta_{jj^\prime} - S_{jl,j^\prime l^\prime} \vert^2
\end{eqnarray}
In this equation, $Y_{l^\prime m_{l^\prime}}(\theta)$ are the spherical harmonics, $M$ is the projection of $J$, 
$k_j=\sqrt{2\mu E_r/\hbar}=\sqrt{2\mu(E-\epsilon_j)/\hbar}$ is the wavenumber in the entrance channel with energy $\epsilon_j$, $E_r$ is the collision energy, and $E$ is the total (kinetic plus rotor) energy.

For the special case $\theta=0$, $Y_{l^\prime m_{l^\prime}}(0)=\sqrt{\frac{2l^\prime+1}{4\pi}} \delta_{m_{l^\prime} 0}$ and the forward scattering amplitude becomes
\begin{eqnarray}  \label{scatt_ampl_0}
f_{jm_j\rightarrow j^\prime m_j}(E_r,0) 
&& = \frac{1}{2 k_j} \sum_{J} \sum_{ll^\prime} i^{l-l^\prime} \sqrt{(2l+1)(2l^\prime +1)} \nonumber \\
&& \times (-)^{l+l^\prime+j+j^\prime} (2J+1) 
\left(\begin{array}{ccc} j & l & J \\ m_j & 0 & -M\end{array}\right) \nonumber \\
&& \times
\left(\begin{array}{ccc} j^\prime & l^\prime & J \\ m_{j} & 0 & -M\end{array}\right) 
\vert \delta_{ll^\prime}\delta_{jj^\prime} - S_{jl,j^\prime l^\prime} \vert^2
\end{eqnarray}

In matter wave experiments, the detected particles have the same translational momentum as the projectiles. Therefore, we only need to compute the elastic scattering amplitude
\begin{equation}
f_{\text{el}, j}(E_r,0)=\frac{1}{2j+1} \sum_{m_j} f_{jm_j\rightarrow j m_j}(E_r,0) \ ,
\end{equation}
which is related to the total collision cross section via the optical theorem
\begin{equation}
\sigma_{\text{tot},j}(E_r)=\frac{4\pi}{k_r}\text{Im}[f_{\text{el},j}(E_r,0)] \ .
\end{equation}

The dependence of the scattering amplitude on the rotational quantum number $j$ must also be taken into account. If we assume a thermal distribution of rotational levels, the amplitude entering Eq. (\ref{refractive_index1}) is given by
\begin{equation}\label{ampl_average_rot}
f(k_r)=\sum_{j} p_j(T) f_{\text{el},j}(k_r)
\end{equation}
where $p_j(T)$ is the population of the rotational level $j$ with energy $\epsilon_j$
\begin{equation}\label{}
p_j(T) = \frac{(2j+1)\exp(-\epsilon_j/kT)}{\sum_{j{^\prime}} (2j{^\prime}+1)\exp(-\epsilon{_j{^\prime}}/kT)}
\end{equation}

To obtain $\rho(v_p)$ for a wide range of projectile velocities and temperatures, we computed the scattering amplitude for relative energies between $10^{-8}$ and 10$^4$ cm$^{-1}$. At low energy, the CC equations were solved numerically \cite{Tscherbul2011b} using a log-derivative algorithm \cite{Manolopoulos1986} on a radial grid from $R_{\min} =3.5 a_0$ to $R_{\max} = 150 a_0$ for all total angular momenta $J$ until convergence of the sum in Eq. (\ref{scatt_ampl_0}). At high energy, we used the coupled states (CS) approximation \cite{McGuire1974} as implemented in the non-reactive scattering code {\small MOLSCAT} \cite{Molscat}. The coupled states method leads to a drastic reduction in computational time and is particularly useful at high energy. Above 1500 cm$^{-1}$, the agreement between the elastic cross sections calculated via the CC or CS methods was within 1\%.

\subsection{Potential energy surface of NaN$_2$}

We adopted the 2D potential energy surface (PES) calculated {\it ab initio} by Loreau {\it et. al.} and presented in Ref. \cite{Loreau2011a}. The potential $V(R,\theta)$ was obtained using the coupled-cluster method and a large basis set, with the N-N bond fixed to its equilibrium distance $r_e=2.0743 a_0$. The potential has a global minimum of 26.92 cm$^{-1}$ at a distance $R_e=10.47 a_0$ between Na and the center of mass of N$_2$ and at an angle $\theta_e=45.0^\circ$. The potential has a saddle point for $\theta=90^\circ$ which lies about 4 cm$^{-1}$ above the global minimum.
The long range part of the potential was constructed assuming a $C_6/R^6$ form and using calculated dispersion coefficients \cite{Loreau2011a}, and was joined smoothly to the {\it ab initio} points using a switching function.
To perform scattering calculations, this potential was expanded in a basis of Legendre polynomials, $V(R,\theta) = \sum_\lambda V_\lambda(R)P_\lambda(\cos\theta)$, retaining terms up to $\lambda=12$. Additional terms did not affect the cross sections or scattering amplitudes.

\section{Results and discussion}\label{section_results}

\subsection{Comparison with the experiment}

\begin{figure}[htp]
\includegraphics[angle=-90,width=.45\textwidth]{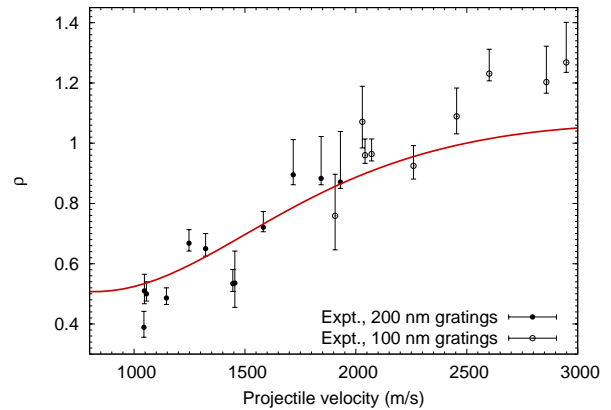}
\caption{Comparison between our calculations (full line) and the experimental values (using 200 nm or 100 nm gratings \cite{Roberts2002}) of $\rho$ as a function of the projectile velocity $v_p$. The temperature is $T=300$K.}
\label{rho_comp_exp}
\end{figure}

The parameter $\rho(v_p)$, defined by Eq. (\ref{rho_def}), is shown in Fig. \ref{rho_comp_exp} for projectile velocities between 800 and 3000 ms$^{-1}$, together with the experimental values from Ref. \cite{Roberts2002}. These experimental data were obtained using 200 nm gratings at low velocities and 100 nm gratings at higher velocities.
We obtain very good agreement between our calculations and experiment for projectile velocities up to around 2000 ms$^{-1}$. 
However, the theoretical values of $\rho$ are lower for velocities higher than 2500 ms$^{-1}$.
We have no explanation for the discrepancy but our calculations show that it is not due to the contribution of inelastic rotational transitions. It is also unlikely to be due to uncertainties in the calculated potential energy surface, despite the sensitivity of $\rho$ to it \cite{Forrey1996}. Discrepancies also occur in the simpler case of sodium matter waves traveling through noble gases \cite{Cronin2009}.

\begin{figure}[htp]
\includegraphics[angle=-90,width=.45\textwidth]{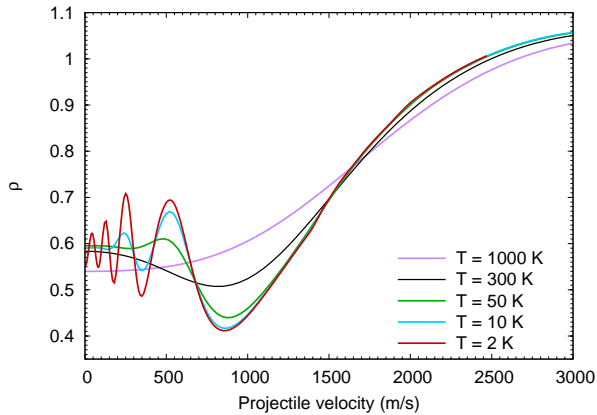}
\caption{Behavior of the ratio $\rho(v_p)$ for various temperatures. }
\label{rho_T}
\end{figure}

In contrast to the noble gases, the experimental data and our calculations for N$_2$ show no sign of glory oscillations. The absence of glory oscillations in the refractive index despite their presence in the Na--N$_2$ elastic collision cross section \cite{Loreau2011a} occurs because they are suppressed by the averaging over the thermal distribution, as found for other systems \cite{Audouard1995,Forrey1997,Bjorgen2011}.
The averaging depends on the temperature as well as on the mass of the target, and the damping would be weaker for heavier particles or for lower temperatures. 
However, even with a temperature lower than 300 K, glory oscillations still would not be observable for the range of velocities used in the experiment. This is illustrated in Fig. \ref{rho_T}, which displays $\rho(v_p)$ for several temperatures and for $v_p$ between 0 and 3000 ms$^{-1}$. It can be seen that $\rho$ does not present any glory oscillations for projectile velocities between 1000 and 3000 ms$^{-1}$, even at low temperature. However, for $v_p$ below 1000 ms$^{-1}$, the oscillations start to appear and become more marked as the temperature decreases. 

\begin{figure}[htp]
\includegraphics[width=.45\textwidth]{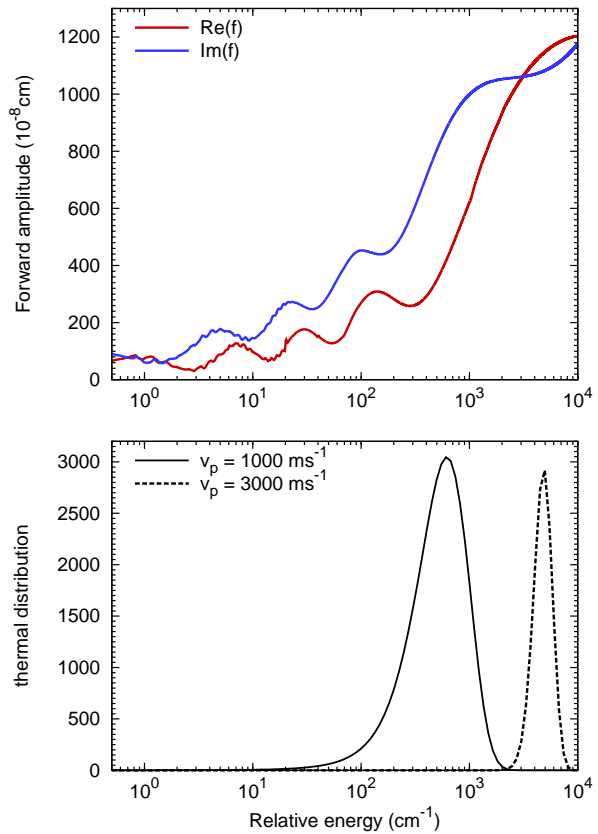}
\caption{Top panel: real and imaginary parts of the forward scattering amplitude $f(k)$ as a function of the relative energy for $j=7$. Bottom panel: distribution function (\ref{MB_distrib2}) as a function of the relative energy for $T=300$ K.}
\label{fig_ampl}
\end{figure}

This can be understood by looking directly at the forward scattering amplitude, shown in the top panel of Fig. \ref{fig_ampl} for energies between 0.5 and $10^4$ cm$^{-1}$. Glory oscillations are indeed present in both the real and imaginary parts of the amplitude, but they are restricted to low energy. The scattering amplitude is shown for the rotational level $j=7$, which is the most populated at $T=300$ K. The amplitude is almost independent of the rotational level over the energy range considered, as already noted for the scattering cross sections \cite{Loreau2011a}. The $j-$averaging (\ref{ampl_average_rot}) has therefore no effect on the value of $\rho$ under the experimental conditions. 

The distribution (\ref{MB_distrib2}) is shown in the bottom panel of Fig. \ref{fig_ampl} for the lowest and highest projectile velocities used in the experiment. These distributions probe the scattering amplitude in the region where no glory oscillations are present, so that the averaging of $f(k_r)$ does not yield any oscillations in $\rho(v_p)$. If we decrease $T$ and $v_p$, it becomes possible to scan the glory oscillations that are present at lower collision energy. At the lowest temperature shown in Fig. \ref{rho_T} ($T=2$K), the thermal distribution (\ref{MB_distrib2}) is so narrow that the features of the scattering amplitude are directly reflected in $\rho(v_p)$; in particular, each maximum (minimum) of $\rho$ corresponds to a maximum (minimum) in the scattering amplitude.

The scattering amplitude also shows a resonance structure at energies below the depth of the potential (27 cm$^{-1}$), which has been discussed in Ref. \cite{Loreau2011a}. This structure depends on the rotational level $j$.

Finally, we also calculated $\rho(v_p)$ using Eq. (\ref{index_Champenois}) instead of Eq. (\ref{refractive_index1}). We found that the two formulas give results that agree within a few percent, even at low projectile velocities.

\subsection{Molecular matter waves}

An interesting question for matter wave interferometry involving molecules is whether exchanging the projectile and target (i.e., using a beam of molecules traveling through a cell of atoms) can provide new insights, in particular related to the determination of properties of the interaction potential. For example, it has already been pointed out in the case of atom interferometry that the magnitude and sign of the scattering length can be obtained by measuring $\rho$ \cite{Vigue1995,Forrey1997}. 
Apart from the difference in mass, which modifies the thermal distribution (\ref{MB_distrib2}), the most obvious consequence of using molecules as projectiles is that if the molecules can be prepared in a definite rovibrational level, then an interferometry experiment could lead to insights into the nature of the interaction potential involving those specific levels. This would be particularly interesting at low temperature as $\rho$ is then expected to be strongly dependent on $v$ and $j$. 

In the following, we will use the scattering amplitudes calculated in the previous section to investigate the behavior of $\rho$ at low projectile velocity and temperature. 
We assume that a beam of nitrogen molecules prepared in a given rotational state travels through a gas cell of sodium atoms at a temperature $T=10^{-8}$ K, so as to avoid any effect of the thermal averaging. As a consequence, the two formulas (\ref{index_Champenois}) and (\ref{refractive_index1}) give the same value for $\rho$. 

In the s-wave regime, the scattering amplitude can be written in term of the phase shift $\delta_0$ as
\begin{equation}\label{swave_ampl}
f(k_r)=\frac{\exp(2i\delta_0)-1}{2ik_r} \approx \frac{\delta_0}{k_r} = -a
\end{equation}
since $\delta_0=-k_ra$ in the limit $k_r\rightarrow 0$, where $a$ is the scattering length.
In the case of atom-molecule collisions, the molecule may be in a ro-vibrationally excited state and the scattering length can be complex, $a=\alpha-i\beta$. The imaginary part $\beta$ is related to the inelastic cross section by $\beta=k_r \sigma_{\text{inel}}/4\pi$ in the limit $k_r \rightarrow 0$ \cite{Balakrishnan1997a}, and the elastic cross section is given by $\sigma_{\text{el}}=4\pi(\alpha^2 + \beta^2)$. Using Eq. (\ref{swave_ampl}), we see that the s-wave amplitude is given in the limit $k_r\rightarrow 0$ by $f=-\alpha+i\beta$, so that $\rho=-\alpha/\beta$.

\begin{figure}[htp]
\includegraphics[angle=-90,width=.45\textwidth]{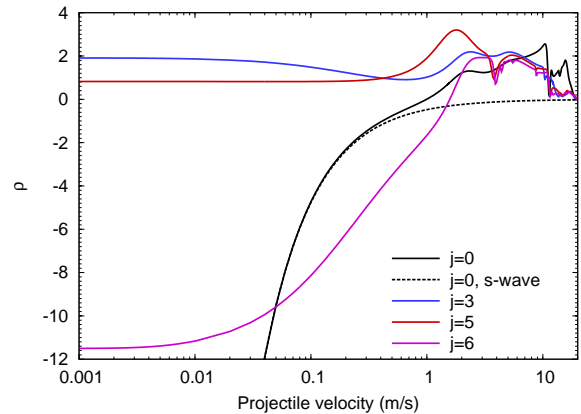}
\caption{Behavior of $\rho(v_p)$ for various rotational levels at low projectile velocity and a temperature $T=10^{-8}$ K. The dashed line corresponds to Eq. (\ref{rho_j0}).}
\label{rho_swave}
\end{figure}

In the case of Na-N$_2$ scattering, $\beta=0$ for $j=0$ and $j=1$ so for these levels $\rho$ will tend to $\pm\infty$ when $k_r\rightarrow 0$. On the other hand, for $j>1$ quenching to levels $j^\prime < j$ is possible and $\rho$ will be constant. Its value will depend on the relative contribution of elastic and inelastic scattering to the total cross section. 
We computed the ratio $\rho$ at low velocity and temperature for various rotational levels. The results for $j=0,3,5$ and 6 are illustrated in Fig. \ref{rho_swave}. 
For $j=0$, $\alpha$ is positive and $\rho$ decreases to $-\infty$, as expected. For $j=3$ or $j=5$, $\alpha$ is negative and has a similar magnitude as $\beta$ so that $\rho\approx 1$, while for $j=6$ $\alpha$ is positive and much greater than $\beta$. 
The behavior of $\rho$ can be understood from Eq. (\ref{swave_ampl}) by retaining higher order terms in the expansion of $\exp(2i\delta_0$). For example, keeping second order terms in the $k$-expansion of the scattering amplitude, we find that the real part of the scattering amplitude for $j=0$ is still Re($f)=-\alpha$, while the imaginary part is Im($f)=k_r\alpha^2$. Therefore, for this level $\rho$ can be approximated at low energy by \cite{Note1}
\begin{equation}\label{rho_j0}
\rho\approx -\frac{1}{k_r\alpha}
\end{equation}
This function is shown in dashed line in Fig. \ref{rho_swave} and agrees very well with the full calculation up to $v_p \sim 1$ ms$^{-1}$.

It should also be noted that at projectile velocities $v_p >20$ ms$^{-1}$, the value of $\rho$ becomes almost independent of the rotational level $j$, as indicated previously. Therefore, it is clear that at high temperatures and projectile velocities, the Na-N$_2$ system does not represent a good experimental case to extract information about the interaction potential from matter wave interferometry. Other systems such as Na$_2$-He (recently studied by Bj\"orgen {\it et al.} \cite{Bjorgen2011}) could be be more promising as $\rho$ seems to have a dependence on the rotational level at higher projectile velocities.

\section{Conclusion}

We computed the index of refraction of sodium matter waves propagating through a molecular nitrogen gas. We obtained the scattering amplitude by performing fully quantal calculations using an accurate {\it ab initio} potential energy surface for the NaN$_2$ complex.
The value of the ratio $\rho$ of the real to the imaginary part of the scattering amplitude agrees well with the experiment except at the highest experimental projectile velocities. We showed that no glory oscillations were observed in the experiment due to the fact that both the temperature and the sodium velocity were too high. 
Finally, we discussed the behavior of $\rho$ in the ultracold regime and its dependence on the rotational level of N$_2$. 
We showed that for ultracold collisions the refractive index is very sensitive to the values of the molecular rotational quantum numbers. This sensitivity may be used for an accurate determination of parameters of the $j-$dependent interaction potentials in atom-molecule systems.

 \acknowledgments
We thank T. Tscherbul for his help with computational issues. 
This work was supported by the U.S. Department of Energy.

\end{document}